\newcommand{\beq}{\begin{equation}}
\newcommand{\eeq}{\end{equation}}
\newcommand{\beqa}{\begin{eqnarray}}
\newcommand{\eeqa}{\end{eqnarray}}
\documentstyle[aps,prl,preprint,amsfonts,epsf]{revtex}
\begin{document}
\def\dfrac#1#2{{\displaystyle{#1\over#2}}}


\preprint{LA-UR-94-1944}

\title{Impurity-Induced Virtual Bound States in $d$-Wave Superconductors
}

\author{A.V.Balatsky$^a$, M.I.Salkola, and A.Rosengren$^b$}
\address{Theoretical Division, Los Alamos National Laboratory, Los Alamos,
 New Mexico 87545}

\date{July 25, 1994}
\maketitle

\begin{abstract}
It is shown that a single, strongly scattering impurity produces a
bound or a virtual bound quasiparticle state inside the gap in a
$d$-wave superconductor. The explicit form of the bound state wave function
is found to decay exponentially with angle-dependent range. These states
provide a natural explanation of the second Cu NMR rate arising from
the sites close to Zn impurities in the cuprates. Finally, for finite
concentration of impurities in a $d$-wave superconductor, we reexamine
the growth of these states into an impurity band, and discuss
the Mott criterion for this band.


\end{abstract}


\newpage


Effects of impurities on the properties of superconductors have been
investigated in great detail for low-temperature \cite{low},
heavy-fermion \cite{heavy}, and high-temperature superconductors
\cite{high}. The main reason for the interest in the effects of
impurities on the superconducting state is the fact that the
superconducting properties are qualitatively modified by impurity
atoms, depending, for example, whether they are magnetic or
non-magnetic.  In principle, this observation can be useful as a
method of identifying the nature of the pairing state in superconductors.
For example, any magnetic impurity will be a strong pair-breaker for
($s$-wave, $d$-wave, etc.) spin-singlet superconductors, in accord with
the generalized Anderson theorem. On the other hand, even scalar
(non-magnetic) impurities are pair-breakers for
``higher-orbital-momentum'' states, such as a $d$-wave pairing state.

The two main approaches in understanding the effects of impurities in
conventional ($s$-wave) superconductors rely either on the strong- or
on the weak-scattering limit. (${\bf a}$) The Abrikosov-Gor'kov (AG
hereafter) formalism \cite{ag} treats impurities in the Born
approximation.  Any impurity problem is characterized by two physical
parameters: the phase shift $\delta_0$ due to impurity scattering
(which we assume to be $s$-wave) and the concentration of impurities
$n_{\rm imp}$. In the AG approach, the only parameter entering the
formalism is the scattering rate, $\tau^{-1}=(2n_{\rm imp}/\pi N_0)
\sin^2\delta_0$, proportional to the product of the concentration and
$\sin^2\delta_0$.  Here $N_0$ is the normal-state density of states at
the Fermi energy, and $\delta_0=N_0V$ is the $s$-wave phase shift for a
weak impurity potential $V$. Therefore, in the limit of dilute
concentration of strong magnetic impurities, the AG approach will yield
a small average scattering rate \cite{com5}. On the other hand, the local
properties of the superconductor near an impurity site, such as the
local density of states and the gap amplitude, will be modified
dramatically. In this limit (${\bf b}$) the Yu-Shiba approach \cite{lu,shiba}
should be used, which treats magnetic impurities in the unitary-scattering
limit with the $s$-wave phase shift $\delta_0 \simeq \pi/2$.  It was shown
by Yu and Shiba that, in the unitary limit, a localized magnetic impurity,
interacting with the spin density of conduction electrons at the impurity site,
produces a true bound state inside the energy gap,
$|\omega|<\Delta_0$, where the density of states vanishes.
Note that, in general, the overlap with the particle-hole
continuum only allows virtual states to be formed with finite lifetime.
The relation between this approach and the AG treatment was established
in \cite{shiba}, where it was shown that in the Born limit one recovers
the AG results, and the bound state is indistinguishably close to the
band edge.

Our work is partially motivated by the fact that non-magnetic impurities
are known to be strong pair-breakers in a nontrivial superconductor.
The Zn substitutions
in cuprates are one example of this.  Although Zn ions are nominally
non-magnetic, $T_c$ is strongly suppressed by Zn substitution of Cu
in the planes \cite{kit,hotta}. Therefore, it is reasonable to assume
that Zn ions are non-magnetic unitary scatterers.

The purpose of this Letter is to address the question of virtual
impurity-bound states in a $d$-wave superconductor and, within this framework,
to explore possible implications of the assumption that the pairing
in cuprates is in the $d_{x^2-y^2}$ channel. We shall generalize
the original Yu-Shiba \cite{lu,shiba} approach to the case of
arbitrary-strength of non-magnetic impurities in a $d_{x^2-y^2}$
superconductor. The results, summarized below, can be easily
generalized for any nontrivial pairing state and may be  relevant
for heavy-fermion superconductors with impurities as well.

Our main results are as follows: (${\bf i}$) A strongly-scattering scalar
impurity is a requirement for a localized, virtual or marginally real,
bound state to exist in a $d$-wave superconductor. It is intuitively
obvious that any strong enough pair-breaking impurity --- magnetic or
 non-magnetic --- will induce such a state.  Indeed, the low-lying
quasiparticle states close to the nodes in the energy gap will be
strongly influenced even by a non-magnetic impurity potential,
resulting in a well-defined bound state in the unitary limit. This should
be compared to the fact that, in $s$-wave superconductors, both
magnetic and resonant non-magnetic impurities produce bound states
inside the energy gap  \cite{ms}. (${\bf ii}$) The energy $\Omega'$ and
the decay rate $\Omega''$ of this state are given by
\beq
\Omega \equiv \Omega'+i\Omega''=
\Delta_0 {\pi c/2 \over {\log(8/\pi c)}} \left[ 1 + {i\pi \over {2}}
{1\over {\log (8/\pi c)}} \right]
\label{eq:1}
\eeq
where $c=\cot \delta_0$. We have assumed impurity scattering to be
close enough to the unitary limit so that the result can be computed to
logarithmic accuracy with $\log (8/\pi c) \gg 1$. It is only in this
limit that the bound state is well-defined. In the unitary
limit, defined as $\delta_0 \rightarrow \pi/2$ ($c\rightarrow 0$), the
virtual bound state becomes a marginally bound one at
$\Omega\rightarrow 0$ with $ \Omega''/ \Omega' \rightarrow 0$. In the
opposite case of weak scattering with $c\lesssim 1$,
the energy of the virtual bound state formally approaches $\Omega' \sim
\Delta_0$ and the state is ill-defined because $\Omega'' \sim \Omega'$
(see Fig.~1).
The wave function of the bound state is found to decay
exponentially, except along the directions of the vanishing gap. (${\bf
iii}$) For a finite impurity concentration $n_{\rm imp}$, we recover
the results obtained earlier for non-conventional superconductors
\cite{heavy,high}. While generally one finds that an impurity band is
formed after averaging over impurity positions, the Mott criterion for
the formation of the impurity band should be modified because of highly
anisotropic impurity states; we suggest that the metal-insulator
transition can be observed in the impurity band.

{\it Single-Impurity Problem.} Consider the single scalar impurity problem
with $H_{\rm int} = \sum_{\vec{k}\vec{k}'\sigma} V
c^\dagger_{\vec{k}\sigma}c_{\vec{k}'\sigma}$,
where $V$ is the strength of the scalar impurity potential at $\vec{r}=0$,
resulting in $s$-wave phase shift $\delta_0$.

 The scattering of quasiparticles off the impurity is described by a
$T$-matrix, $\hat{T}(\omega)$, which is independent of wavevector. The
Green's function in the presence of an impurity is
$\hat{G}_{\vec{k}\vec{k}'}(\omega)=
\hat{G}^{(0)}_{\vec{k}}(\omega)\delta_{\vec{k}\vec{k}'} +
\hat{G}^{(0)}_{\vec{k}}(\omega) \hat{T}(\omega)
\hat{G}^{(0)}_{\vec{k}'}(\omega)$, where both
$\hat{G}^{(0)}_{\vec{k}}(\omega)$ and $\hat{T}(\omega)$ are matrices in
Nambu space. Here $[ \hat{G}^{(0)}_{\vec{k}}(\omega)]^{-1} = \omega
\hat{\tau}_0 - \Delta_{\vec{k}} \hat{\tau}_1 - \xi_{\vec{k}}
\hat{\tau}_3$, where $\xi_{\vec{k}}$ is the quasiparticle energy,
$\Delta_{\vec{k}}= \Delta_0 \cos 2\varphi$ is the gap function with
$d_{x^2-y^2}$-symmetry, $\hat{\tau}_i$ ($i=1,2,3$) are the Pauli
matrices, and $\hat{\tau}_0$ is the unit matrix in Nambu
particle-hole-spinor space.

 From the previous analysis \cite{heavy,ag,shiba}, it is known that
$\hat{T} = T_0 \hat{\tau}_0$ for $s$-wave scattering and a particle-hole
symmetric system. Therefore, the only relevant term in the $T$-matrix takes the
form
$
T_0(\omega) = G_0(\omega)/[c^2 - G_0(\omega)^2],
$
where $G_0(\omega) = {1\over 2\pi N_0} \sum_{\vec{k}} {\rm Tr} \,
\hat{G}^{(0)}_{\vec{k}}(\omega)
\hat{\tau}_0$.
The virtual and bound states in the single-impurity problem are given by
the poles of the $T$-matrix
\beq
c= \pm G_0(\Omega), \label{eq:2}
\eeq
which is an implicit equation for $\Omega$ as a function
of $c$, the strength of impurity scattering. The two signs in Eq.~(\ref{eq:2})
are a result of the particle-hole symmetry. Choosing the gap function at the
Fermi surface so that $\Delta(\varphi)= \Delta_0 \cos 2\varphi$, one finds
$G_0(\omega)=\left\langle  \omega/
\sqrt{\Delta(\varphi)^2 -\omega^2}\right\rangle _{FS}$, where the
angular brackets denote averaging over the Fermi surface; for simplicity,
we take $\langle   \bullet  \rangle _{FS} = \int \bullet \ d\varphi/2\pi$
\cite{com2}. For $|\omega|\ll \Delta_0$,
one finds
\beq
G_0(\omega) = {2\omega\over \pi \Delta_0} \left( \log {4\Delta_0\over \omega}
- {i\pi\over 2}\right). \label{eq:3}
\eeq
In principle, the solution of Eq.~(\ref{eq:2}) is complex, indicating
a resonant nature of the quasiparticle state, better described as a virtual
state. This is easily seen form Eq.~(\ref{eq:1}),
which solves Eq.~(\ref{eq:2}) to logarithmic accuracy. However, as
$c\rightarrow0$,
the resonance can be made arbitrarily sharp. For $c=0$, the virtual state
becomes
a marginally well-defined state bound to the impurity.  Exact numerical
solution of Eq.~(\ref{eq:2}) as a function of $c$ is shown in Fig.~2.  To our
knowledge, this result has not been claimed before. As $c\rightarrow 1^-$,
$\Omega'$ and $\Omega''$ increase without bound so that
$\Omega''/\Omega' \rightarrow 1^-$, and the solution becomes unphysical.
For $c>1$, no solution has been found for $\Omega$ \cite{h}.

There are important physical implications of these bound states in a $d$-wave
superconductor. Consider the most interesting case of unitary impurities
in the dilute limit, separated by a distance greater than the coherence length
$\xi$.
Before averaging over impurities, these bound states are {\it nearly localized}
close to the impurity sites (see below) and can substantially modify the local
characteristics of
the superconductor: for example, the local density of states and the local
NMR relaxation rates of atoms close to the impurities.

Consider a local density of states, defined as $N({{\vec {r}}},\omega)
= -{1\over{\pi}}{\rm Im}\,G({\vec {r}},{\vec {r}};\omega+i0^+)$, with
the total Green's function in the presence of the impurity
$\hat{G}(\vec{r},\vec{r}\,';\omega)=
\hat{G}^{(0)}(\vec{r}-\vec{r}\,',\omega) +
\hat{G}^{(0)}(\vec{r},\omega) \hat{T}(\omega)
\hat{G}^{(0)}(\vec{r}\,',\omega)$, the second term describing the local
distortion due to the impurity. Using the eigenstates representation of
$G({\vec {r}},{\vec {r}\,'};\omega) = \sum_{n}\langle \psi_n^*({\vec
{r}}) \psi_n({\vec {r}\,'})\rangle/(\omega - E_n)$, we find two
terms in the local density of states $N({{\vec {r}}},\omega) =
N(\omega) + N_{{\rm imp}}({\vec {r}},\omega) \simeq \sum_{n}\langle
\psi_n^*({\vec {r}}) \psi_n({\vec {r}})\rangle \delta(\omega - E_n)$,
for well defined states. The first term originates from the bulk
quasiparticles, which are described by plane-wave eigenstates with
$E_{{\vec{k}}} = \sqrt{\xi^2_{{\vec{k}}} + \Delta^2_{{\vec{k}}}}, \ \
G^{(0)}({\vec {r}},\omega) =
\sum_{{\vec{k}}} [u_{\vec{k}}^2/(\omega - E_{{\vec{k}}}) +
v_{\vec{k}}^2/(\omega + E_{{\vec{k}}}) ]$, where
$u_{\vec{k}}$ and $v_{\vec{k}}$ are the standard Bogoliubov factors.
The bulk density of states is constant in the system with
$N(\omega)/N_0 = \omega/\Delta_0$, for $\omega \ll \Delta_0$. The
second term, $N_{{\rm imp}}( {\vec {r}},\omega) = -{1\over{\pi}} {\rm
Im} \, [{\hat{G}^{(0)}({\vec{r}},\omega) \hat{T}(\omega)
\hat{G}^{(0)}({\vec{r}},\omega)}]_{11} $, originates from the virtual
quasiparticle state created at the impurity: $N_{{\rm imp}} ({\vec
{r}},\omega) =- {1\over{\pi}}{\rm Im}\sum_{n} \langle\psi_{{\rm imp},
n}^*({\vec {r}}) \psi_{{\rm imp}, n}({\vec {r}})\rangle/(\omega -
E_{{\rm imp},n}+i0^+)$.  As an important example, consider the limit
of unitary scattering for which the resonant state is formed at
$E_{{\rm imp},n}\equiv\Omega\rightarrow 0$.  Because ${\rm Im}\,
G^{(0)}({\vec{r}},\omega = 0) =-\pi N(\omega = 0) = 0$, only the imaginary part
of the T-matrix contributes to $N_{{\rm imp}}$ and the bound-state probability
density is found to decay as the inverse second power of the
distance from the impurity along the nodes of the gap function,
\begin{mathletters}
\beq
N_{{\rm imp}}({\vec {r}},\omega = 0) =
{\rm Re}\,[\hat{G}^{(0)}({\vec{r}},\omega = 0)]^{2} \propto  r^{-2},
\label{eq:20}
\eeq
and exponentially in the vicinity of the extrema of the gap function,
\beq
N_{{\rm imp}}({\vec {r}},\omega = 0) \propto [\xi(\varphi)/r]^{-1}
  e^{-2r/\xi(\varphi)},
\eeq
\end{mathletters}
where $\xi(\varphi)$ is the angle-dependent coherence length of the
superconductor, naturally defined as $\xi(\varphi) = \hbar
v_F/|\Delta(\varphi)|$. The fact that the impurity state is marginally
bound is reflected in the logarithmically divergent normalization.
This divergence should be cut off at an average distance between impurities at
any finite concentration.  More generally, for an arbitrary position of the
resonance, taking into account that only one state has been produced
with $ E_{{\rm imp},n} = \Omega' +i\Omega''$, we find
$
N_{{\rm imp}} ({\vec {r}},\omega) = {1\over{\pi}}\sum_i
F({\vec {r}} - {\vec {r_i}})\, \Omega_i''/[(\omega-\Omega_i')^2 +
\Omega_i''^2],
$
where we have introduced the sum over different impurities,
located at ${\vec {r_i}}$, and $F({\vec {r}}-{\vec {r_i}}) =
\langle\psi_{\rm imp}^*({\vec {r}}-{\vec {r_i}})
\psi_{\rm imp}({\vec {r}}-{\vec {r_i}})\rangle$ is
the probability density of the $i$-th impurity state.

The local variations of the density of states can be probed directly,
in principle, by scanning-tunneling microscopy. However the NMR
experiments on Cu in Zn-doped cuprates are quite revealing as well. From
Eq.~(\ref{eq:20}) and below, one concludes immediately that the local NMR
signal would show two distinct relaxation rates (or even the hierarchy
of rates): one coming from the Cu
sites, far away from the impurities, and another from the
sites, close to the impurities. The Cu sites near the impurities will
be sensitive to the higher local  density of states and will have a
higher relaxation rate at low temperatures \cite{com15}.  At  finite
impurity concentration ($\sim 2\%$), the volume-averaged density of
states will have a finite limit at $\omega\rightarrow 0$, as follows from
Eq.~(\ref{eq:20}).  The
relaxation rates of  Cu atoms close to and away from an impurity will,
therefore, have the same temperature dependence $(T_1T)^{-1} = const$,
but will be of a different magnitude.

 This behaviour has been observed experimentally:  Ishida  {\it et al.}
\cite{kit}  have measured two NMR relaxation rates for Cu in Zn-doped
YBa$_2$Cu$_3$O$_{7-\delta}$. The second NMR signal with higher
relaxation was inferred arising from the near-impurity Cu sites.
A direct comparison of our prediction for local quantities, as probed by
NMR, will require a specific model and is left for a future
publication.

We would like to contrast our picture of the dilute limit of strongly
scattering centers to the usual approach of averaging over impurities
at finite concentration. If one considers averaging over impurities,
two NMR relaxation rates, arising from unequivalent sites,
cannot  be resolved; the local inhomogeneous aspect of the localized
states is lost after averaging over impurity positions.

For practical purposes the
distinction between the impurity bound states and continuum in our case
is not as well defined as in $s$-wave superconductors. Any finite
temperature will produce a finite lifetime for these bound states, and
they will be hybridized with the continuum of low-energy
quasiparticles.

{\it Finite Concentration of Impurities.}  Consider the growth of the
impurity band with finite concentration of strongly scattering
impurities.   As was mentioned above, scalar (non-magnetic) impurities
are pair-breakers for  any nonconventional superconductor, and they
substantially change the low-energy spectrum of superconducting
quasiparticles. This problem has been addressed earlier in great detail
(for example, see \cite{heavy,high}). Here we will repeat the main
steps and give results for the quasiparticle scattering rate and
low-energy density of states for completeness.

For finite impurity concentration, the self-consistent Green's function,
averaged over impurity positions, obeys the Dyson equation $ \langle
\hat{G}_{\vec{k}}(\omega)\rangle ^{-1} =
\langle \hat{G}^{(0)}_{\vec{k}}(\omega)\rangle ^{-1} - \hat{\Sigma}(\omega)$
with $\hat{\Sigma}(\omega) = n_{\rm imp}\hat{T}(\omega)$ in the single-site
approximation. The $T$-matrix has to be determined self-consistently with
$\langle G_0(\omega)\rangle  = {1\over 2\pi N_0} \sum_{\vec{k}} {\rm Tr}
\,\langle
\hat{G}_{\vec{k}}(\omega)\rangle \hat{\tau}_0$.
As above [see the remark preceding Eq.(\ref{eq:2})], only $\Sigma_0\equiv
{1\over 2} {\rm Tr}\, \hat{\Sigma}\hat{\tau}_0$
is nonzero. The algebra is straightforward, and for unitary scattering
yields
\beq
\gamma \simeq   \sqrt{n_{\rm imp}(\Delta_0
/\pi N_0)},
\label{eq:9}
\eeq
where $\gamma = - {\rm Im} \,\Sigma(\omega\rightarrow 0)$ is the
scattering rate for low-energy quasiparticles. For $\omega \lesssim
\gamma$, the density of states  is determined by impurities and is
finite:   $N_{\rm imp}(0)/N_0 = 2\gamma/\pi\Delta_0$. The characteristic
width of the impurity-dominated region is $\omega^* \simeq \gamma
\propto \sqrt{n_{\rm imp}}$.

The origin of the finite density of states   is the impurity band,
grown from the impurity-induced bound states (consider $c = 0$).  Scaling of
the impurity bandwidth $\gamma \propto
\sqrt{n_{\rm imp}}$ has been obtained earlier for the case of
paramagnetic impurities in an $s$-wave superconductor \cite{shiba}.
The fact that $\gamma \propto \sqrt{n_{\rm imp}}$ is obeyed in the case
of a $d$-wave superconductor with scalar impurities as well supports
our claim that the low-energy states in a disordered $d$-wave
superconductor are indeed formed from the bound states at finite
concentration.

Here we shall comment on the condition for the formation of the
impurity band with extended states, i.e., the Mott criterion. Consider
the case of a (3D) $s$-wave superconductor. A magnetic impurity generates
a bound state with the impurity wave function $|\Psi_{\rm imp}| \sim
\exp(-\alpha r)$, $\alpha \sim \xi^{-1}$ \cite{lu,shiba}. For the true
conduction band to be formed, the overlap between localized states
should be large enough. This leads to the Mott criterion for the minimum
concentration $n_{\rm imp}^{1/3} \alpha^{-1} \geq 0.2$ \cite{Mott}. In
practice, for conventional superconductors this implies a very low
critical  concentration $n^c_{\rm imp} \sim (a/\xi)^3$ ($a$ being the
lattice constant), as the coherence length $\xi$ is relatively large
\cite{com35}.  The situation is qualitatively different in  a $d$-wave
superconductor. The wave function of impurities has lobes sticking out
in the directions of vanishing gap, i.e., when $\cos2\varphi = 0$ [see
Eq.~(\ref{eq:20})], and the overlap between impurities is larger along
these directions. The Mott transition in a system of strongly
anisotropic  localized states is an interesting problem and it should
be expected that the Mott criterion should be modified with a somewhat
lower constant in the r.h.s.\ of $n_{\rm imp}^{1/2} \xi = const$ and it
follows that $n^{c}_{\rm imp} \propto (a/\xi)^2 \sim 1\%$, which is
within the current experimental range. A short coherence length of
high-$T_c$ superconductors ($\xi \sim 20 {\rm \AA}$) suggests the possibility
of the experimental observation of the metal-insulator transition in the
impurity band in a $d$-wave superconductor (assuming that localization
effects are small).

{\it In conclusion}, we find that a strongly scattering potential impurity
produces a resonant or a marginally bound state inside the gap in a $d$-wave
superconductor. The wave function of the impurity bound state is highly
anisotropic with $1/r$ decay along the nodes of the gap and exponential
with angle-dependent decay range  otherwise. These bound states change
the local density of states $N_{\rm imp}(r, \omega)$ dramatically, which could
be probed experimentally, e.g., in NMR. The short coherence length of
high-$T_c$ superconductors may lead to a finite critical
concentration $n^c_{\rm imp}$ within the range of a few percents and allow
a direct observation of the metal-insulator transition in the
impurity band.

\

We are grateful to P.~Littlewood and M. Takigawa for discussions, to
P.~Hirschfeld for explaining us the results of Ref. \cite{h},
and to D.~Rabson for useful comments on the manuscript. We would also like
thank L.~Yu for his interest and bringing Ref.~\cite{lu} to our attention.
This work was supported by the U.S.\ Department of Energy and by the Swedish
Natural Research Council.


\figure{\noindent FIG.1 \ The relative density of states $N(\omega)/N_0$ for
the $d$-wave superconductor in the absence of impurities (thin solid line),
the impurity-induced bound state for $c\rightarrow 0$ (thick solid line),
and the virtual bound state for $c>0$ (dashed line); $N_0$ is the normal-state
density of states at the Fermi energy. The finite lifetime ($\Omega''\ne0$)
of the virtual bound state in the $d$-wave
superconductor results from the finite density of states $N(\omega)
\propto \omega$ for small $\omega$ from nodal quasiparticles, in contrast
to the true bound state in the $s$-wave
superconductor. Exactly at $\omega=0$, the density of quasiparticle states
is zero and the virtual bound state becomes marginally bound on the edge
of the particle-hole continuum.
}

\figure{\noindent FIG.2 \ The energy $\Omega=\Omega'+i\Omega''$ of the
virtual bound state in the one-impurity problem, given by Eq.~(\ref{eq:2}),
as a function of impurity strength $c$: the shown quantities are $\Omega'$
(solid line), $\Omega''$ (dashed line), and $\Omega''/\Omega'$
(chain-dashed line). A spherical Fermi surface and
$\Delta_{\vec{k}}=\Delta_0\cos 2\varphi$ have been assumed. The width
$\Omega''$
of the virtual bound state is always smaller than its energy $\Omega'$ in
the neighbourhood of unitary scattering. In contrast, for weak scattering
$c\sim 1$, $\Omega''\sim \Omega'$ and an impurity-induced virtual state does
not exist. The inset shows a comparison between the exact result and
the asymptotic approximation (dotted lines), as computed to
logarithmic accuracy by Eq.~(\ref{eq:1}) for $\Omega$. }


\begin{thebibliography}{99}
\bibitem{com2} We assume that the energy gap has line nodes in three dimensions
with weak quasiparticle dispersion along the $z$ axis; an extension to a
general three-dimensional case is straightforward.
\bibitem{com15} The change of the matrix elements and antiferromagnetic
correlations  will  also result in the change of the normal state relaxation
rate for  Cu sites  close to the impurity. These effects are, we believe,  less
important at low temperatures, where the drastic increase of the local density
of states will dominate and lead to an increased relaxation rate at sites near
the impurity.
\bibitem{com35} For the experimentally relevant concentrations $n_{\rm imp}
\simeq 0.5 - 1 \%$, the impurity band with ``metallic'' conductivity  is
well established.
\end{thebibliography}
\end{document}